\documentclass{ws-ijmpcs}

\begin{document}

\markboth{Bisnovatyi-Kogan G.S., Krivosheev Yu.M.}
{Thermal balance of the jet in the microquasar SS433}

%
\catchline{}{}{}{}{}
%

\title{THERMAL BALANCE OF THE JET IN THE MICROQUASAR SS433 }

\author{BISNOVATYI-KOGAN G.S., KRIVOSHEEV YU.M.}

\address{Space Research Institute Rus. Acad. Sci., Profsoyuznaya 84/32,\\ Moscow, 117997, Russia\\
gkogan@iki.rssi.ru; krivosheev@iki.rssi.ru}

\maketitle

\begin{history}
\received{Day Month Year}
\revised{Day Month Year}
\end{history}

\begin{abstract}
Thermal balance of the jet in the source SS433 is considered with account of radiative and adiabatic cooling, and different heating mechanisms. We consider jet heating by the inverse Compton effect of coronal hard X-ray quanta on jet electrons, the influence of shock wave propagation along the jet, and jet kinetic energy transformation into heat via Coulomb collisions of jet and corona protons.  The most important heating mechanism for the source SS433 turns out to be Coulomb collisions of jet particles with the surrounding medium.

\end{abstract}

\ccode{PACS numbers: 11.25.Hf, 123.1K}

\section{Introduction}	

SS433 is a unique massive X-ray binary system with precessing relativistic jets. It is situated at a distance of approximately $5 \,\,{\rm kpc}\approx 1.5\times 10^{22}$ cm, nearly in the galactic plane. The optical companion V1343 Aquilae was first identified in the survey of stars exhibiting $H_\alpha$ (656 nm) emission by Stephenson and Sanduleak \cite{ss77}.
It is one of the
brightest stars in the Galaxy, the bolometric luminosity of the
object assuming isotropic radiation is $L_{bol}\sim10^{40}$ erg/s
\cite{Cher82}. SS433 is a close binary system with an orbital
period of 13.1 days \cite{Cher81}. The uniqueness of this source
comes from existence of narrow oppositely directed subrelativistic
jets, which emit red and blue-shifted, periodically variable lines.
A commonly accepted model of this object suggest that a continuous
regime of supercritical accretion of gas onto the relativistic star
is maintained. Thus, a supercritical accretion disk forms, together
with narrow  jets of gas propagating perpendicular to the disk plane
from the central regions of the disk and having the relativistic
speed of 0.26c. The optical star fills its critical Roche lobe,
providing a powerful and almost continuous flow of gas into the
region of the relativistic star at a rate of
$\sim10^{-4}M_{\odot}$/yr \cite{fab04}.

The
INTEGRAL observations of SS433 discovered X-ray emission in the
range  3 to 90 keV, which remains to be visible over the whole
period of the binary. That leads to suggestion of the presence of an
extended hot region (corona) in the central parts of the accretion disk
\cite{Cher05}. Modeling of this
spectrum by Monte-Carlo simulations \cite{kriv09} provided an information about the physical parameters
of this system. Here we analyze a thermal balance of the jet, and investigate sources of its heating, needed to maintain
the temperature of the jet on the level, imposed by the X-ray and optical observations. The detailed description will be published in \cite{bkk11}.

\section{Monte-Carlo modeling}

\begin{figure}
\centerline{\psfig{file=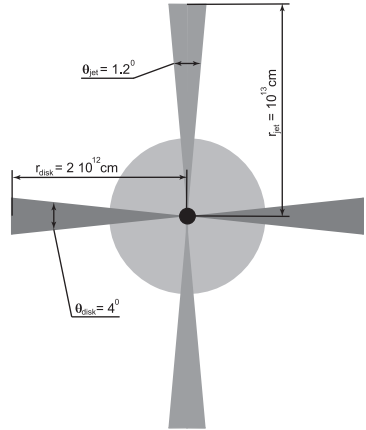,width=4.7cm}}
\vspace*{8pt}
\caption{Schematic picture of the source SS433}
\label{fig1}
\end{figure}

We accept, that this binary system consists of an optical star and a black hole, surrounded by an accretion disk, and a hot corona, with a couple of jets. We have
considered the accretion disk within the radius
$1.5\cdot 10^{12}$ cm. Far from the central parts of the disk standard
accretion disk theory \cite{ss} can be used to estimate its
geometrical thickness $h=r\tan \frac{\theta_{disk}}{2}$, and the disk opening angle  $\theta_{disk}\approx 2$ degrees. As follows from observations,  the jets have a conical shape with the opening angle $\theta_{jet}\approx 1.2$ degrees. All parameters are listed in the  Fig.\ref{fig1} from \cite{kriv09}, $r_0=10^{11}$ cm.

\begin{figure}
\centerline{\psfig{file=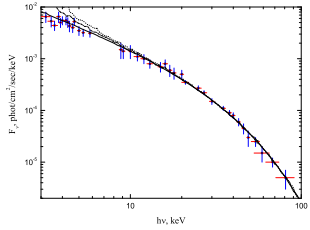,width=4.7cm}}
\vspace*{8pt}
\caption{The results of the simulations for different values of accretion disk effective temperature: $6.0\cdot10^{5}$~K (solid line), $9.4\cdot10^{5}$~K (dash-dotted line), $1.1\cdot10^{6}$~K (dotted line). The best fit occurs when $T_{ef}=6.0\cdot10^{5}$~K. $\tau_{cor}=0.24$, $T_{cor}=19$~keV, $r_{cor}=6.4\cdot10^{11}$~cm, $\dot{M_{jet}}=4\cdot10^{19}$~g/s.}
\label{fitdisk}
\end{figure}

The density profile $n \sim r^{-2}$, and the adiabatic temperature profile $T\sim \varrho^{2/3}\sim r^{-4/3}$,
 have been accepted. It was obtained \cite{kriv09} that the best coincidence of the model with observational takes place at the adiabatic temperature distribution along the jet.
In the figure Fig.\ref{fitdisk} the best fit SS433 spectrum in the range from 3 to 90 keV is presented. The spectrum corresponds to precessional moment, when the angle between jet axis and the line of sight is equal 60 degrees and the disk is maximally 'face-on'.

\section{Thermal balance of the jet}

The equation for the energy balance of the matter in the stationary jet with account of heating, and adiabatic and radiative cooling is written as \cite{bkk11}

\begin{equation}
 \frac{3}{2}\frac{v_{jet}k}{m_{p}}\frac{dT}{dr}
 =\frac{v_{jet}}{m_{p}}kT\frac{1}{\rho}\frac{d\rho}{dr} + q
\label{eneq2}
\end{equation}
Here $\varepsilon$ is the internal energy of the gas unit mass, $q$ includes radiative cooling, and different types of heating,
$p$ is a pressure, $V = 1/\rho$ is a specific gas volume. Bremstrahlung losses (erg/g/s) are determined by the expression, see \cite{bk1}

\begin{equation}
 q^{ff}=\int q^{ff}_{\nu,\theta}d\nu d\theta=
 16\left(\frac{2\pi}{3}\right)^{3/2}
 \frac{e^{6}}{hm_{p}m_{e}c^{2}}n_{jet}\sqrt{\frac{kT}{m_{e}c^{2}}}.
= 0,852\cdot10^{-3}n_{jet}\sqrt{T}.
\label{ff}
\end{equation}
Solution of (\ref{eneq2}) for $q=q^{ff}$, obtained in \cite{KovSh89}, have shown a very rapid temperature drop. The value of $T$ drops formally to zero at $r\approx 1.5 r_0$, what is in contradiction with  observational data. In \cite{bkk11} all energy losses have been used in calculations, including bound-free, and bound-bound transitions from \cite{kir}.

Jet heating by Compton interaction with hot photon gas emitted by the corona and the disk was investigated, and it was shown, that it is much less than the radiation losses \cite{bkk11}.

\subsection{Heating by shock waves}

\begin{figure}
\centerline{\psfig{file=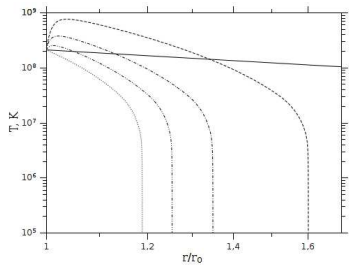,width=4.7cm}}
\vspace*{8pt}
\caption{The temperature profiles along the jet with account of heating by the shock waves, and full radiative losses. The dash line corresponds to $M$=2, dash-dot line corresponds to $M$=3, dash-dot-dot line corresponds to $M$=5. The dot line corresponds to the case without shock heating, and the full line corresponds to adiabatic expansion,  $r_0=10^{11}$ sm.}
\label{figsw}
\end{figure}

The shock wave generation may happen during the jet propagation through the surrounding gas. It was shown in \cite{bkk11} that due to the entropy increase in the shock, the  average heating rate due to shock dissipation is determined as

\begin{equation}
\label{qsw}
  q_{sw} \approx 
 \frac{kT}{(\gamma - 1) m_p \Delta t}\left[ \ln \left( \frac{2\gamma
M^2}{\gamma + 1}
 - \frac{\gamma -1}{\gamma + 1} \right) + \gamma \ln \left( \frac{\gamma -
1}{\gamma + 1}
 + \frac{2}{M^2 (\gamma + 1)} \right) \right]
\end{equation}
the value of $\Delta t$=1 sec was taken equal to the characteristic time of
SS433 variability in the X-rays \cite{fab04}.
 The Mach number is decreasing due to transformation of the kinetic flow energy into heat in the shock.. The time averaged equation for the  Mach number change along the jet is written as

 \begin{equation}
\label{eq1}
 \frac{2(M^2-1)}{(\gamma + 1)^2}\rho_{0} c_{s}^{3}
 \left[
 \left(3+\frac{1}{M^2} \right)
  \frac{dM}{dr} + \frac{1}{M}(M^2- 1)
  \left(\frac{1}{\rho_0}\frac{d\rho_0}{dr}+\frac{3}{2T}\frac{dT}{dr} \right)
  \right]
\end{equation}
$$= -\rho_{0} \frac{(\gamma+1)M^2}{2+(\gamma-1)M^2}q_{sw}.
$$
The temperature distribution with account of shock heating and radiative losses is written as

\begin{equation}
\label{eq2}
 \frac{1}{\gamma-1}\frac{dT}{dr}=\frac{T}{\rho_{0}}
 \frac{d\rho_{0}}{dr}+\frac{m_p}{k}\left(\frac{q_{sw}}{M c_s}-\frac{q_{br}}{v_{jet}}\right).
\end{equation}
The solution of (\ref{eq1}),(\ref{eq2}), for the shock formation at the disk origin, is shown in Fig.\ref{figsw}. It may be seen that the shock is rapidly dissipating. The shock heating could be important only if the shocks are originating in the jet all along its length \cite{bkk11}.

\subsection{Heating by Coulomb collisions}

Transformation of the jet kinetic energy into the heat happens when the protons from the corona and circumstellar gas are entering the jet. During collisions  these protons are thermalized, transforming their relative kinetic energy into the heat, decreasing the kinetic energy of the jet.
Maximal heating happens when the proton entering the jet loose its whole, and remains in the jet. In this case the rate of heating is determined as 
$ q_{coll,max}=v_{jet}^{2}(v_{th}/2d_{jet})
 (n_{cor}/n_{jet}) \sim r^{-1}.
$
In reality, the proton may go through the jet, loosing only part of its energy. For the jet with above mentioned parameter coulomb collisions with protons of the jet may extract only small part of the energy because of small cross-section, at which the mean free path of the proton in the jet greatly exceeds its radius. Introducing $\tau_{coll}$, equal to a ration of the proton energy lost in the jet, we have \cite{bkk11}

\begin{equation}
\label{tau_Coul}
 \tau_{coll}=n_{jet}d_{jet}\frac{2\int d\sigma
 \Delta E}{m_{p}v_{jet}^2}\approx 2\cdot 10^{-4}, \,\,\, {\rm at}\,\,\,
  \frac{1}{m_pv_{jet}^2/2}\int\Delta E d\sigma=2\pi\rho_{p}^{2}
  \ln\left(\frac{\lambda_{D}}{\rho_p}\right).
\end{equation}
Here $\rho_{p}=e^{2}/m_{p}v_{jet}^{2}\approx 2.1\cdot 10^{-15}$~ñì,
$r=6.4\cdot10^{11}$~ñì, $n_{jet}=n_{jet,0}/41\approx 2\cdot 10^{13}$~cm$^{-3}$, $d_{jet}=2r\tan{(\theta_{jet}/2)}\approx 1.3\cdot 10^{10}$~ñì,
 $\ln\left(\frac{\lambda_{D}}{\rho_p}\right)\approx 25$. It follows that the jet heating due to coulomb collisions is much smaller than the radiative cooling $\frac{q_{coll}}{q_{br}}\approx 6\cdot10^{-3}$.

A presence even of a small magnetic field changes the situation. When a giroradius $r_L$ of the proton is smaller than the radius of the disk the proton may spend larger time inside the jet and can loose a larger portion of its energy.
The equality $r_L=r_{jet}$ is reached at $B_{jet} = \frac{m_p c v_{th}}{e r_{jet}}\approx 1.3\cdot 10^{-4}$ Gs, at the jet  origin. In reality the magnetic field is probably much larger, so we may suggest that the proton is loosing a considerable part of its energy inside the jet. Here we try to find what part of the energy the proton should loose inside the inner, X-ray jet,  to obtain the temperature distribution close to the adiabatic one.

\begin{figure}
\centerline{\psfig{file=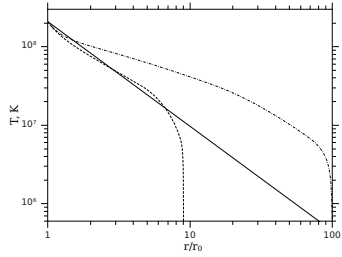,width=4.7cm}}
\vspace*{8pt}
\caption{
 Temperature profile in the jet with account of the radiative cooling, and heating by the coulomb collisions, with
  $a=(r/r_{0})^{-3/2}$ (dash line), $a=(r/r_{0})^{-1.163}$ (dash-dot line).
  The full line corresponds to pure adiabatic cooling.}
 \label{Tcoll2}
\end{figure}

We choose the heating by collisions in the form
$
q_{coll}(r)=q_{coll,max}(r)*a(r).
$
Several examples for the temperature dependence at different choice $a(r$ are given in Fig.\ref{Tcoll2}.
The case with $a=(r/r_{0})^{-1.163}$ gives a good coincidence with the adiabatic curve up to $10 r_0$, where the main source of the radiation is situated. It is clear that the whole adiabatic curve may be fitted with a more complicated function $a=r)$.

\begin{figure}
\centerline{\psfig{file=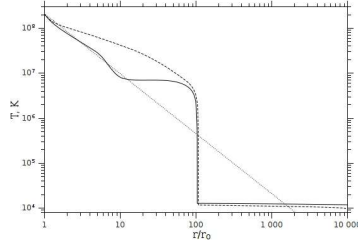,width=4.7cm}}
\vspace*{8pt}
\caption{ Temperature profile in the jet with account of the radiative cooling, and heating by the coulomb collisions, with  $a(r)\sim r^{-1.163}, r<r_{cor};
 a(r)\sim r^{-0.48}, r>r_{cor}$ (dash line);
the full line corresponds to the analytic function $a(r)$, which gives the best fit to the adiabatic curve in the X-ay jet, and constant temperature optical jet; dot line corresponds to a pure adiabatic expansion.}
 \label{optjet}
\end{figure}

The optical jet in SS433 is situated at distances
$r\sim10^{14-15}$~cm   ($10^{3-4}r_{0}$), and has almost constant temperature $T\sim10^{4}$~Ê \cite{Davidson+}. Because of radiative losses, and adiabatic expansion, a heating is necessary to support the constancy of the temperature.  We consider the same mechanism of the optical jet heating. The protons come from the circumstellar gas with adiabatically decreasing temperature $T\sim \rho^{2/3}$ outside the outer radius of the corona  $r=r_{cor}=6,4\cdot10^{11}$~cm. Inside the corona the constant temperature is accepted  $T_{0}=2.2\cdot10^{8}$~K. The density profile is taken the same inside and outside the corona $n\sim r^{-2}$, with the the density at the bottom of the corona $n_{cor,0}=4.3\cdot10^{12}$~cm$^{-3}$. It was obtained in \cite{bkk11}, that the temperature profile all over the jet may be reproduced by the choice of $a(r)$, see Fig.\ref{optjet}  from
\cite{bkk11}.

\section{Conclusions}

The most effective heating mechanism of the jet in SS 433 is connected with kinetic energy losses by collisions with surrounding matter (protons), in presence of a very moderate magnetic field. Collisions with protons from circumstellar gas may support $T\sim 10^4$ Ê in the optical jet. Shocks distributed over the jet may give an input into heating.
The losses of the jet kinetic energy to its heating are relatively very small $~10^{-4}$, and the velocity change along the jet is hardly observable.

\section*{Acknowledgments}

This work  was partially supported by the RFBR grant 11-02-00602, the RAN programm P20, and RF President Grant NSh-3458.2010.2.

\end{document}